\documentclass[aps,prl,twocolumn,showpacs,groupedaddress]{revtex4}  
\usepackage{graphicx}  
\usepackage{dcolumn}   
\usepackage{bm}        
\usepackage{amssymb}   

\begin{document}


\hspace{5.2in} \mbox{Fermilab-Pub-07/488-E}

\title{Measurement of the $p\bar{p}\rightarrow WZ + X$ cross section 
at $\sqrt{s}=1.96$~TeV and limits on $WWZ$ trilinear gauge couplings
}

%
\author{V.M.~Abazov$^{35}$}
\author{B.~Abbott$^{75}$}
\author{M.~Abolins$^{65}$}
\author{B.S.~Acharya$^{28}$}
\author{M.~Adams$^{51}$}
\author{T.~Adams$^{49}$}
\author{E.~Aguilo$^{5}$}
\author{S.H.~Ahn$^{30}$}
\author{M.~Ahsan$^{59}$}
\author{G.D.~Alexeev$^{35}$}
% ...

\affiliation{\vspace{0.1 in}(The D\O\ Collaboration)\vspace{0.1 in}}
% ...
\affiliation{$^{5}$University of Alberta, Edmonton, Alberta, Canada,
                Simon Fraser University, Burnaby, British Columbia, Canada,
                York University, Toronto, Ontario, Canada, and
                McGill University, Montreal, Quebec, Canada}
% ...
\affiliation{$^{28}$Tata Institute of Fundamental Research, Mumbai, India}
% ...
\affiliation{$^{30}$Korea Detector Laboratory, Korea University, Seoul, Korea}
% ...
\affiliation{$^{35}$Joint Institute for Nuclear Research, Dubna, Russia}
% ...
\affiliation{$^{49}$Florida State University, Tallahassee, Florida 32306, USA}
% ...
\affiliation{$^{51}$University of Illinois at Chicago,
                Chicago, Illinois 60607, USA}
% ...
\affiliation{$^{59}$Kansas State University, Manhattan, Kansas 66506, USA}
% ...
\affiliation{$^{65}$Michigan State University,
                East Lansing, Michigan 48824, USA}
% ...
\affiliation{$^{75}$University of Oklahoma, Norman, Oklahoma 73019, USA}
% ...

\date{September 18, 2007}

\begin{abstract}
We present measurements of the process 
$p\bar{p} \rightarrow WZ+X \rightarrow \ell^{\prime} 
\nu_{\ell^{\prime}} \ell \bar{\ell}$ at $\sqrt{s}=1.96$ TeV, 
where $\ell$ and $\ell^{\prime}$ are electrons or muons. 
Using 1 fb$^{-1}$ of data from the D0  experiment,  we observe $13$
candidates with an expected background of $4.5\pm0.6$ events and  
measure a cross section $\sigma(WZ)=2.7^{+1.7}_{-1.3}$ pb.
From the number of observed events and the 
$Z$ boson transverse momentum distribution, 
we limit the trilinear  $WWZ$ gauge
couplings to 
$-0.17 \le \lambda_Z \le 0.21$ $(\Delta \kappa_Z = 0)$
at the 95\% C.L. for a form factor scale $\Lambda=2$ TeV. Further,
assuming that $\Delta g^Z_1 = \Delta\kappa_Z$, we find 
$-0.12 \le \Delta\kappa_Z \le 0.29$ $(\lambda_Z=0)$ at the 95\% C.L.
These are the most restrictive limits on the $WWZ$ couplings available
to date.
\end{abstract}

\pacs{14.70.Fm, 13.40.Em, 13.87.Ce, 14.70.Hp} 
\maketitle 


The $SU(2)_L\times U(1)_Y$ structure of the standard model (SM) Lagrangian
requires that the massive electroweak gauge bosons, the $W$ and $Z$ bosons, 
interact with one another at trilinear and quadrilinear vertices.
In the SM, the production cross section for $p\bar{p}\rightarrow WZ+X$, 
$\sigma(WZ)$, depends on the strength of the $WWZ$ coupling, 
$g_{WWZ} = -e \cot \theta_{W}$, 
where $e$ is the positron charge and $\theta_{W}$ is the weak mixing angle. 
At $\sqrt{s}=1.96$ TeV, the SM predicts $\sigma_{WZ}=3.68\pm 0.25$  
pb~\cite{ref:RunIITheorySigma}.
Any significant deviation from this prediction would be evidence for new 
physics.  

The $WWZ$ interaction can be parameterized by a generalized effective 
Lagrangian~\cite{ref:HPZH,ref:HISZ} with $CP$-conserving
trilinear gauge coupling parameters 
(TGCs) $g^Z_1$, $\kappa_Z$, and $\lambda_Z$ that
describe the coupling strengths of the vector bosons to the weak field. 
The TGCs are commonly presented as deviations from their SM values, 
i.e. as $\Delta g^Z_1 = g^Z_1 - 1$, $\Delta \kappa_Z = \kappa_Z - 1$, and 
$\lambda_Z$, where $\lambda_Z = 0$ in the SM. 
Since tree-level unitarity restricts the anomalous couplings
   to their SM values at asymptotically high energies, each of the 
   couplings must be parameterized as a form factor, e.g. 
   $\lambda_Z(\hat{s})=\lambda_Z/(1+\hat{s}/\Lambda^2)^2]$, where 
   $\Lambda$ is the form factor scale and $\hat{s}$ is the 
   square of the invariant mass of the $WZ$ system.
New physics will result in anomalous TGCs and an enhancement in the production 
cross section as well as modifications to the shapes of 
kinematic distributions, such as the $W$ and $Z$ bosons transverse momenta.  
Because the Fermilab Tevatron is the only particle accelerator that can 
produce the charged state $WZ+X$, this measurement provides a unique 
opportunity to study the $WWZ$ TGCs without any assumption on the 
values of the $WW\gamma$ couplings. 
Measurements of TGCs using the $WW$ final 
state~\cite{CDF1, run1wz, LEPTGC, D0RunIIWW, cdfwwwz2007} are sensitive to 
both the $WW\gamma$ and $WWZ$ couplings at the same time and
must make some assumption as to how they are related to each 
other.

$WZ$ production measurements and studies of $WWZ$ couplings have been
presented previously. The D0  Collaboration
measured  $\sigma_{WZ} = 4.5 ^{+3.8}_{-2.6}$ 
pb, with a 95\% C.L. upper limit
of 13.3 pb, using 0.3 fb$^{-1}$ of $p\bar{p}$ collisions at $\sqrt{s}=1.96$
TeV~\cite{ref:RunIIWZ300pb}. The observed number of candidates was used to 
derive the most restrictive available limits on anomalous $WWZ$ couplings.
More recently, the CDF Collaboration measured
$\sigma_{WZ} = 5.0 ^{+1.8}_{-1.6}$ pb using 1.1 fb$^{-1}$ of $p\bar{p}$
collisions at $\sqrt{s}=1.96$ TeV~\cite{ref:CDFRunIIWZPub},
but did not present any results on $WWZ$ couplings.

This communication
describes a significant improvement to the previous D0  analysis.
Not only is the data sample more than three times larger, but an
improved technique is used to constrain the $WWZ$ couplings.
Instead of merely the total number of observed events,
the number and the $p_T$ distribution of the $Z$ bosons $(p_T^Z)$
produced in the collisions are compared to the expectations of 
non-SM $WWZ$ couplings,
significantly increasing the power of the $WWZ$ coupling measurement
over previous measurements~\cite{run1wz,ref:RunIIWZ300pb}.

We search for $WZ$ candidate events in final states with three charged leptons, 
referred to as trileptons, produced when $Z\rightarrow\ell^+\ell^-$ 
and $W\rightarrow\ell'\nu$, where $\ell$ and $\ell '$ are 
$e^{\pm}$ or $\mu^{\pm}$.
SM backgrounds can be suppressed by requiring three isolated high-$p_T$ 
leptons and large missing transverse energy 
 (\mbox{${\hbox{$E$\kern-0.6em\lower-.1ex\hbox{/}}}_T$})
from the neutrino.  
The combined branching fraction for these four possible final 
states ($eee,~ee\mu,~\mu\mu e$ and $\mu\mu\mu$) is $1.5\%$ ~\cite{ref:pdg}.

D0 is a multipurpose detector~\cite{ref:run2det}
composed of several subdetectors and a fast triggering system. 
At the center of the detector is a central tracking system, consisting of a
silicon microstrip tracker (SMT) and a central fiber tracker (CFT),
both located within a 2~T superconducting solenoidal
magnet. These detectors are optimized for tracking and
vertexing at pseudorapidities~\cite{ref:pseudo} 
$|\eta|<3$ and $|\eta|<2.5$, respectively.
The liquid-argon and uranium calorimeter has a
central section (CC) covering $|\eta| < 1.1$, 
and two end calorimeters (EC) that extend coverage
to $|\eta|\approx 4.2$, with all three housed in separate
cryostats~\cite{ref:run1det}. An outer muon system, covering $|\eta|<2$,
consists of a layer of tracking detectors and scintillation trigger
counters in front of 1.8~T iron toroids, followed by two similar layers
after the toroids~\cite{ref:run2muon}.  

Electrons are identified by their distinctive pattern of energy deposits in 
the calorimeter and by the presence of a track in the central tracker that 
can be extrapolated from the interaction vertex to a cluster of energy in the 
calorimeter.
Electrons measured in the CC (EC) must have $|\eta|<1.1$ $(1.5<|\eta|<2.5)$.  
Electrons must have transverse energy $E_T > 15$ GeV and 
be isolated from other energy clusters.  
A likelihood variable, formed from the quality of the electron track and 
its spatial and momentum match to the calorimeter cluster 
and the calorimeter cluster information, is used to discriminate 
electron candidates from instrumental backgrounds.

Muons tracks are reconstructed using information from the muon drift chambers and
scintillation detectors and must have a matching central track with 
$p_T >15$ GeV/$c$.  Candidate muons are 
required to be isolated in the calorimeter and tracker to 
minimize the contribution of muons originating from jets~\cite{topprd}.


Events collected from 2002--2006 using single muon, single electron, 
di-electron, and jet triggers were used for signal and background studies.
The integrated luminosities~\cite{lum} for the $eee$, $ee\mu$,
$\mu\mu e$, and $\mu\mu\mu$ final states are $1070$ pb$^{-1}$,
$1020$ pb$^{-1}$, $944$ pb$^{-1}$, and $944$ pb$^{-1}$, respectively.
There is a common $6.1\%$ systematic uncertainty on the integrated 
luminosities.


The $WZ$ event selection requires three reconstructed, well-isolated
leptons with $p_T > 15$ GeV/$c$. All three leptons must be associated
with isolated tracks that originate from the same collision point
and must satisfy the electron or muon identification criteria outlined
above. To select $Z$ bosons, and further suppress background, the 
invariant mass of a like-flavor lepton pair must fall within the range $71$ to
$111$ GeV/$c^2$ for $Z\rightarrow ee$ events, and $50$ to $130$ GeV/$c^2$
for $Z\rightarrow \mu\mu$ events, with the mass ranges set by the
mass resolution. For $eee$ and $\mu\mu\mu$ decay channels, the
lepton pair with invariant mass closest to that of the $Z$ boson mass
are chosen to define the $Z$ boson daughter particles. The 
\mbox{${\hbox{$E$\kern-0.6em\lower-.1ex\hbox{/}}}_T$}
is required to be greater than 20 GeV, consistent
with the decay of a $W$ boson. 
The transverse recoil of the $WZ$ system, calculated using the 
vector sum of the transverse momenta of the charged leptons
and missing transverse energy, is required to be less than 50~GeV/$c$. 
This selection reduces the background contribution from $t\bar{t}$ 
production to a negligible level.

%
%

\begin{figure}
\includegraphics[scale=0.4]
{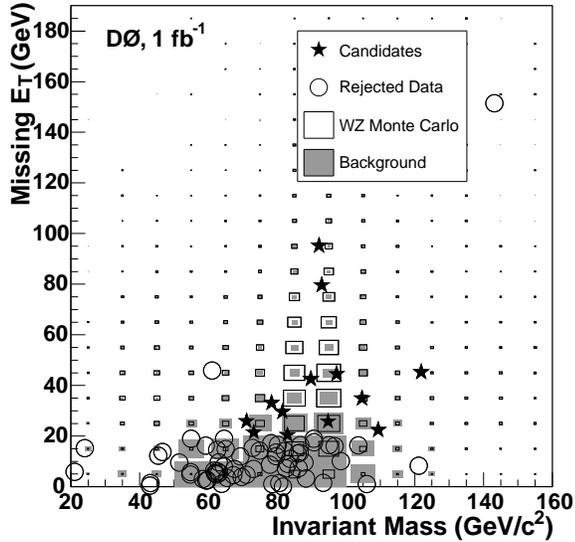}
\caption{\label{Fig:massvsmet}
 \mbox{${\hbox{$E$\kern-0.6em\lower-.1ex\hbox{/}}}_T$}
versus dilepton invariant mass of $WZ$ candidate events.  
The open boxes represent the expected $WZ$ signal. 
The grey boxes represent the sum of the 
estimated backgrounds.  The black stars are the data that survive
all selection criteria. The open circles are data that fail either
the dilepton invariant mass criterion or have 
\mbox{${\hbox{$E$\kern-0.6em\lower-.1ex\hbox{/}}}_T$}
$ < 20$ GeV. }
\end{figure}

\begin{figure}
\includegraphics[angle=0,scale=0.4]{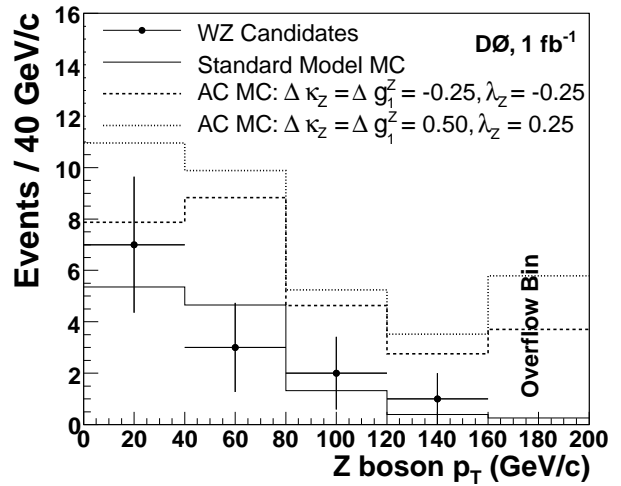}
\caption{\label{Fig:Zpt}The reconstructed $Z$ boson $p_T$ of the 
$WZ$ candidate events used in the $WWZ$ coupling parameter limit setting 
procedure.  The solid histogram is the expected sum of signal and background for 
the case of the $WWZ$ coupling parameters set to their SM values.  The dotted and 
double dotted histograms are the expected sums of signal and background for two 
different cases of anomalous $WWZ$ coupling parameter values.  The black 
dots are the data.  The final bin is the overflow bin.}
\end{figure}

%
%
$WZ$ event detection efficiencies are determined for each 
 of the four final states. Monte Carlo (MC) events are generated 
 using {\sc pythia}~\cite{ref:pythia} and a {\sc geant}~\cite{ref:geant} 
 detector simulation and then processed using the same reconstruction 
 chain as the data.
Lepton identification efficiences are determined
from study of $Z$ bosons in the D0 data. 
The average efficiencies for detecting an electron or muon with $E_T$ 
$(p_T) >$  15 GeV are $(91 \pm 2)\%$ and $(90 \pm 2) \%$, 
respectively. The trigger efficiency for events with two (or more) 
electrons is estimated to be $(99 \pm 1)\%$. 
For events with two or three muons, the trigger efficiencies 
are estimated to be $(91\pm5) \%$ and $(98\pm 2)\%$, respectively.
The kinematic and geometric acceptances range from 29\%  
for the $eee$ decay mode to 45\% for the $\mu\mu\mu$ decay mode.
It is also necessary to account for $\tau \rightarrow e,\mu$ 
final states of $WZ$ that contribute to the signal.  
The number of $\tau$ events expected to satisfy the selection criteria is 
$0.67\pm 0.11$ events. These are treated as signal in the cross section 
analysis, but are treated as background in the TGC analysis.
Table~\ref{tab:eventsum} summarizes the efficiency determinations.

%
%
A total of 13 $WZ$ candidate events is found. 
Figure~\ref{Fig:massvsmet} shows 
\mbox{${\hbox{$E$\kern-0.6em\lower-.1ex\hbox{/}}}_T$} versus 
the dilepton invariant mass for the
background, the expected $WZ$ signal, and the data, including
the candidates. Table~\ref{tab:eventsum}
also details the number of candidates in each channel.

The main background for $WZ\rightarrow \ell^\prime \nu \ell \ell$ are $Z+X$ 
events where $X$ is a jet that has been
misidentified as an electron or muon.  
We assess the background from $Z$+jets production by using an 
inclusive jet data sample that is selected with an independent jet 
trigger.  Events characteristic of QCD two-jet production are used 
to measure the probability, as a function of jet $E_T$ and $\eta$,  
that a single jet will be misidentified as a muon or electron.  Next, 
sub-samples of $ee$+jets, $e\mu$+jets, and $\mu\mu$+jets events are 
selected using the same criteria as for 
the $WZ$ signal except that the requirements for a third lepton in the 
event are dropped.  The single jet-lepton misidentification 
probabilities are then convoluted with the measured jet distributions 
in the dilepton+jets sub-samples to provide an estimate of the 
background from $Z$+jets events.  The contribution for all four 
decay modes totals $1.3\pm0.1$ events.

All other backgrounds are determined using MC. Non-negligible backgrounds 
include SM $ZZ$ production, $Z\gamma$ production, and  $W^*Z$, $WZ^*$, 
or $W\gamma^*$ production.  We define these 
processes as three-lepton final states produced through the decay 
of one on-mass-shell and one off-shell vector boson.
These backgrounds and their determination are described as follows.


$ZZ$ production becomes a background when 
both $Z$ bosons decay to charged leptons and one of the final state leptons 
escapes detection, thus mimicking a neutrino.
The total contribution from $ZZ$ production is $0.70\pm 0.08$ events.  

$Z\gamma$ final states can be misidentified as $WZ$ events
if the photon is mis-reconstructed as an electron and there is 
sufficient \mbox{${\hbox{$E$\kern-0.6em\lower-.1ex\hbox{/}}}_T$}.
We estimate the $\ell \bar{\ell} + \gamma$ contribution using 
$Z+\gamma$ MC~\cite{ref:baur} combined with the probability for a 
photon to be misidentified as an electron $(4.2\pm1.5\%)$ determined 
from studies of events with photons. This process is a background 
only to the $eee$ and $\mu \mu e$ final states. The total contribution 
is $1.4\pm 0.5$ events.


The contribution to the background from off-shell bosons 
should be nearly the same as occurs in similar processes and a fraction 
relative to the expected signal is determined from $ZZ$ MC events 
generated using {\sc pythia}.  
It depends on the decay channel and varies from $8\%$ for the $ee\mu$ 
mode to $15\%$ for the $\mu\mu\mu$ mode. The uncertainties 
include all of those used for the signal plus an additional
16\% systematic component to account for uncertainties in the 
off-shell component of the MC.
The total contribution of this background is $0.99\pm 0.19$ events.

To cross check the background estimates, we compare the number 
of observed events with that expected when we do not apply the
dilepton invariant mass selection and the 
\mbox{${\hbox{$E$\kern-0.6em\lower-.1ex\hbox{/}}}_T$}
selection.
We expected to observe $12.5 \pm 1.4$ events from signal and 
$62.9 \pm 8.4$ events from backgrounds. We observe the
$78$ events shown in Fig.~\ref{Fig:massvsmet}.

\begin{table}
\caption{\label{tab:eventsum}
The numbers of candidate events, expected signal events, 
and estimated background events, and the overall detection 
efficiency for the four final states.}
\begin{ruledtabular}
\begin{tabular}{ccccc}
Final & Number of & Expected & Estimated   & Overall \\ 
State & Candidate & Signal   & Background  & Efficiency \\ 
      & Events    & Events   & Events      & \\ \hline
$eee$ & 2         & $2.3\pm 0.2$ 
                             & $1.2\pm 0.1$& $0.16\pm 0.02$ \\ 
$ee\mu$ 
      & 1         & $2.2\pm 0.2$ 
                             & $0.46\pm 0.03$ 
                                           & $0.17\pm 0.02$ \\ 
$\mu\mu e$ 
      & 8         & $2.2\pm 0.3$ 
                             & $2.0\pm 0.4$& $0.17\pm 0.03$ \\ 
$\mu\mu\mu$ 
      & 2         & $2.5\pm 0.4$ 
                             & $0.86\pm 0.06$ 
                                           & $0.21\pm0.03$ \\ \hline
Total & 13        & $9.2\pm 1.0$ 
                             & $4.5\pm 0.6$& -- \\ 
\end{tabular}
\end{ruledtabular}
\end{table}


The SM predicts that $9.2\pm 1.0$ $WZ$ events are 
expected to be observed in the 
data sample.  The probability for the background, 
$4.5\pm 0.6$ events, to 
fluctuate to 13 or more events is $1.2\times 10^{-3}$, 
which translates to a one-sided Gaussian significance of $3.0 \sigma$,  
determined by using a Poisson distribution for the number of observed 
events in each channel convoluted with a Gaussian  to model the 
systematic uncertainty on the background.
 A likelihood method~\cite{ref:stats} taking into account correlations
among systematic uncertainties is used to determine the
most probable $WZ$ cross section. The cross section  
$\sigma(WZ)$  is 
$2.7^{+1.7}_{-1.3}$~pb, where the $\pm 1 \sigma$ uncertainties
are the $68\%$ C.L. limits from the minimum of the negative log likelihood.   
The uncertainty is 
dominated by the statistics of the number of observed events.

\begin{figure}
\includegraphics[scale=0.4]{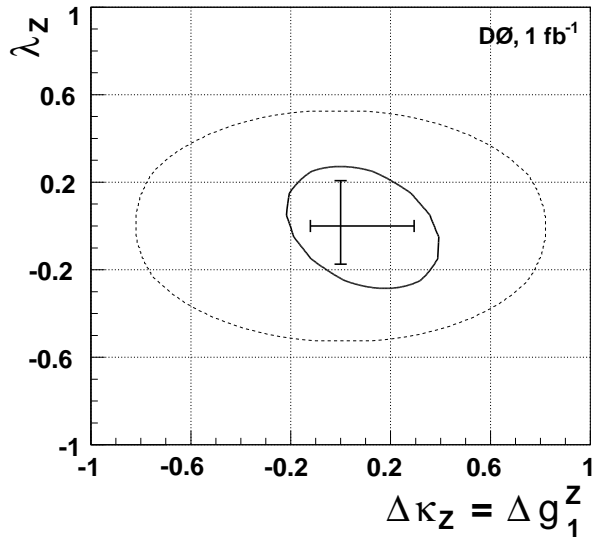}
\caption{\label{fig:twoDcontour} Two-dimensional 95\% C.L. contour limit in 
$\Delta g_1^Z = \Delta \kappa_Z$ versus $\Delta \lambda_Z$ space
 (inner contour).  
The form factor scale for this contour is $\Lambda = 2$ TeV.  
The physically 
allowed region (unitarity limit) is bounded by the outer contour.
The cross hairs are the 95\% C.L. one-dimensional limits.}
\end{figure}

By comparing the measured cross section and $p_T^Z$ distribution 
to models with anomalous 
TGCs, we set one- and two-dimensional limits on the three 
$CP$-conserving coupling parameters.  A comparison of the observed $Z$ boson 
$p_T$ distribution with MC predictions is shown in Fig.~\ref{Fig:Zpt}.  
We use the Hagiwara-Woodside-Zeppenfeld (HWZ)~\cite{ref:HWZ}
leading-order event generator 
processed with a fast detector and event reconstruction simulation to 
produce events with anomalous $WWZ$ couplings and simulate their efficiencies
and acceptances. 
The HWZ event generator does not account for $\tau$ final 
states, and as a result, we treat the 
$0.7$ event $\tau$ contribution as background
for the $WWZ$ coupling limit setting procedure. 
The method used to determine the 
coupling limits is described in Ref.~\cite{ref:Cooke&Illinworth}. 
Limits are set on the coupling parameters 
$\lambda_Z$, $\Delta g^Z_1,\text{ and } \Delta \kappa_Z$. 
Two-dimensional grids are constructed in which the parameters $\lambda_Z$ 
and $\Delta g^Z_1$ are allowed to vary simultaneously. 
Table \ref{Tab:1DLimits} presents the one-dimensional 95\% C.L. 
limits on $\lambda_Z$, $\Delta g^Z_1$ and $\Delta \kappa_Z$.  
Figure~\ref{fig:twoDcontour} presents the two-dimensional 95\% C.L. limits 
under the assumption $\Delta g^Z_1 = \Delta \kappa_Z$~\cite{ref:HISZ}
for $\Lambda=2$ TeV.
The form factor scale, $\Lambda$~\cite{ref:formfactor}, 
associated with each grid, is chosen such that the limits are within 
the unitarity bound.  
\begin{table}
\caption{\label{Tab:1DLimits}One-dimensional 95\% C.L. intervals 
on $\lambda_Z$, $\Delta g^Z_1$, and $\Delta \kappa_Z$ for two sets 
of form factor scale, $\Lambda$.}
\begin{ruledtabular}
\begin{tabular}{cc}
$ \Lambda = 1.5 \text{~TeV} $ & $ \Lambda = 2.0 \text{~TeV} $     \\ \hline
 $ -0.18<\lambda_Z<0.22$       & $ -0.17<\lambda_Z<0.21$          \\
 $ -0.15<\Delta g^Z_1<0.35 $   & $ -0.14< \Delta g^Z_1<0.34 $     \\
 $ -0.14<\Delta \kappa_Z = \Delta g^Z_1 <0.31$ 
                               & $-0.12<\Delta \kappa_Z = 
                                               \Delta g^Z_1 <0.29$ \\ 
\end{tabular}
\end{ruledtabular}
\end{table}


In summary, we present the results of a search for 
$WZ$ production in $1.0$ fb$^{-1}$ of
$p\bar{p}$ collisions at $\sqrt{s} = 1.96$ TeV.  
We observe 13 trilepton candidate events  
with an expected $9.2\pm 1.0$ 
signal events and $4.5\pm 0.6$ events from  
background.  This gives an observed significance of 3.0$\sigma$. 
We measure the $WZ$ production cross section  
to be $2.7^{+1.7}_{-1.3}$~pb, in agreement with the SM prediction.  
We use the measured cross section and $p_T^Z$ 
distribution to improve constraints on $WWZ$ trilinear gauge couplings 
by a factor of two over the previous best results.


%
We thank the staffs at Fermilab and collaborating institutions, 
and acknowledge support from the 
DOE and NSF (USA);
CEA and CNRS/IN2P3 (France);
FASI, Rosatom and RFBR (Russia);
CAPES, CNPq, FAPERJ, FAPESP and FUNDUNESP (Brazil);
DAE and DST (India);
Colciencias (Colombia);
CONACyT (Mexico);
KRF and KOSEF (Korea);
CONICET and UBACyT (Argentina);
FOM (The Netherlands);
Science and Technology Facilities Council (United Kingdom);
MSMT and GACR (Czech Republic);
CRC Program, CFI, NSERC and WestGrid Project (Canada);
BMBF and DFG (Germany);
SFI (Ireland);
The Swedish Research Council (Sweden);
CAS and CNSF (China);
Alexander von Humboldt Foundation;
and the Marie Curie Program.
%


\begin{thebibliography}{99}
\bibitem{ref:pseudo}
     The D0  coordinate system is cylindrical with the 
     $z$-axis along the proton beamline and the polar and azimuthal
     angles denoted as $\theta$ and $\phi$ respectively. The 
     pseudorapidity is defined as $\eta=-\ln{\tan{\theta}/2}$.
\end{thebibliography}
\end{document}